\documentclass[12pt]{iopart}
\usepackage{graphicx}
\usepackage{iopams}

\begin{document}

\title[]{High temperature behavior of Sr-doped layered cobaltites
Y(Ba$_{1-x}$Sr$_{x}$)Co$_2$O$_{5.5}$: phase stability and structural
properties}
\author{G Aurelio$^1$, R D S\'{a}nchez$^{1,2}$, J Curiale$^{1,3}$, G J Cuello$^4$}
\address{$^1$ Consejo Nacional de Investigaciones Cient\'{\i }ficas y
T\'{e}cnicas, Centro At\'{o}mico Bariloche - Comisi\'{o}n Nacional
de Energ\'{\i}a At\'{o}mica, Av. Bustillo 9500, 8400 S. C. de
Bariloche, RN, Argentina}
\address{$^2$ Instituto Balseiro, Universidad Nacional de Cuyo.}
\address{$^3$ CNRS-LPN Laboratoire de Photonique et de
Nanostructures, Route de Nozay 91460 Marcoussis, France.}
\address{$^4$Department of Applied Physics II, Ikerbasque, UPV/EHU, E-48080 Bilbao,
Spain and Institut Laue Langevin, F-38042 Grenoble, France}
\ead{gaurelio@cab.cnea.gov.ar}
\date{\today}

\begin{abstract}
In this article we present a neutron diffraction in-situ study of the
thermal evolution and high--temperature structure of layered
cobaltites Y(Ba, Sr)Co$_2$O$_{5+\delta}$. Neutron
thermodiffractograms and magnetic susceptibility measurements are
reported in the temperature range 20~K~$\leq T \leq$~570~K, as well
as high resolution neutron diffraction experiments at selected
temperatures. Starting from the as-synthesized samples with
$\delta\approx 0.5$, we show that the room temperature phases remain
stable up to 550~K, where they start loosing oxygen and transform to
a vacancy-disordered ``112" structure with tetragonal symmetry. Our results also show how the so-called ``122"
structure can be stabilized at high temperature (around 450~K) in a
sample in which the addition of Sr at the Ba site had suppressed its
formation. In addition, we
present the structural and magnetic
properties of the resulting samples with a
new oxygen content $\delta\approx 0.25$ in the temperature range
20~K~$\leq T \leq$~300~K.

\end{abstract}

 \maketitle

\section{Introduction \label{Introduction}}

The layered cobaltites \emph{R}BaCo$_{2}$O$_{5+\delta}$ (\emph{R}
being a rare earth or Yttrium) are a family of magnetic oxides which
present a very rich variety of phenomena being intensively
studied~\cite{01Aka,05Tas,05Fro,05Pla,05Kha,06Fro,07Kha,08Lue,08Kha,08Fro,08Dia,08Mot}.
The major attraction of this system is that many of its physical
properties may be tuned by the interplay of a number of factors: the
\emph{R} cation size, the oxygen non-stoichiometry affecting the Co
valence state, the vacancies structural order, and -- as we have
shown recently~\cite{07Aur2,09Aur} -- the disorder introduced by
doping with different cations, such as Ca and Sr.

The crystal structure of layered cobaltites is always composed of
different possible combinations of a main lattice related to a basic
perovskite by doubling the cell along the $c$ axis due to the
stacking sequence of
[CoO$_2$]-[BaO]-[CoO$_2$]-[\emph{R}O$_{\delta}$] planes. In the
limiting case $\delta=1$ \cite{05Nak} all Co atoms present an
octahedral oxygen coordination and a mixed valence state 3+/4+. The
oxygen vacancies when $0 \leq \delta < 1$ always locate in the rare
earth planes [RO$_{\delta}$]. Therefore, the structure for
$\delta=0$ consists of all Co atoms in a pyramidal oxygen
coordination~\cite{97Mar,99Mai} with a mixed valence state 2+/3+. In
between, a series of superstructures may form, the most studied one
being the so-called ``122" superstructure, consisting of an ordered
array of 50\% Co atoms in octahedral oxygen coordination and 50\% in
a pyramidal environment and a valence 3+. This array results in a
doubling of the cell also along the $b$ axis, breaking the
tetragonal symmetry found in the other cases, but it only occurs for
$\delta$ values very close to 0.5~\cite{99Aka}.

In a recent study we have shown that when \emph{R}=Y, layered
cobaltites present phase separation and their ground state
corresponds to the coexistence of two distinct antiferromagnetic
(AFM) phases \cite{07Aur2,09Aur}. We arrived at this picture after a
systematic study of the effects of doping with Ca
and Sr, which strongly stabilize each of the two competing phases in
the system: a phase with orthorhombic symmetry which adopts a
ferrimagnetic state with spin-state order (ferri-SSO), and a phase
with tetragonal symmetry which adopts a G-type AFM state. The phase separation scenario has
been confirmed in the last couple of years~\cite{08Lue,08Kha,09Koz,09Rav}, and the
complexity of the magnetism and transport properties still calls for
further investigation.

Although the most intensively studied aspects of these systems remain the low-temperature properties and their attractive underlying fundamental physics, during the last years an interest has also
emerged in the field of cathodes for intermediate temperature solid
oxide fuel
cells~\cite{07Tar,08Tar,08Tar2,09Liu,08Kim,10Zha,10Choi,10Tsv}. The ordered nature of layered cobaltites improves
transport compared to non-ordered perovskites, combining
oxygen transport with good electronic conduction above the
metal-insulator transition temperature.
Recent studies propose that even better candidates for
such applications are the Sr-doped cobaltites
Y(Ba$_{0.5}$Sr$_{0.0}$)Co$_2$O$_{5+\delta}$ due to their excellent
area-specific resistance~\cite{09Kim}.

In this work we extend our study of
cobaltites YBaCo$_2$O$_{5+\delta}$,
Y(Ba$_{0.95}$Sr$_{0.05}$)Co$_2$O$_{5+\delta}$ ($x_{\rm{Sr}}=0.05$)
and Y(Ba$_{0.90}$Sr$_{0.10}$)Co$_2$O$_{5+\delta}$
($x_{\rm{Sr}}=0.10$) turning to their high--temperature behavior. We have used neutron diffraction techniques to
follow the evolution of the different phases in the temperature
range 20~K~$\leq T \leq$~570~K. Along the experiment the oxygen
content was not kept fixed, which gave us the opportunity to study
one same sample with different $\delta$ values. In the following, we describe the sequence of phases observed on warming and cooling in samples with $x_{\rm{Sr}}=0$ and $x_{\rm{Sr}}=0.05$ --allowing for
oxygen loss-- and in sample $x_{\rm{Sr}}=0.10$ avoiding oxygen loss.
Magnetic susceptibility measurements are presented and
discussed in the light of neutron diffraction results and a consistent description is given as a function of dopant concentration, temperature and oxygen content.

\section{Experimental methods \label{experimental}}

The polycrystalline samples used in the present work correspond to
those reported in Refs.~\cite{07Aur2} for $x_{\rm{Sr}}=0$ and
~\cite{09Aur} for $x_{\rm{Sr}}=0.05$ and $x_{\rm{Sr}}=0.10$. Between the previous and present experiments, samples were stored in a dessicator containing silica gel, at room temperature (RT) and ambient pressure.

Neutron diffraction data were obtained in different experiments at
the High Flux Reactor of Institute Laue--Langevin ILL, Grenoble,
France. High--temperature thermodiffraction was performed on the
high--intensity two--axes diffractometer D20. For samples
$x_{\rm{Sr}}=0$ and $x_{\rm{Sr}}=0.05$ we used a wavelength of
$1.31$~{\AA} in the range 300~K~$\leq T \leq$~570~K. Data were
collected on warming and on cooling with programmed ramps of
1.2~K/min. Samples were kept 30 min at 570~K to stabilize the
structures. In a second experiment, a wavelength of $2.42$~{\AA} was
used to study samples $x_{\rm{Sr}}=0.05$ and $x_{\rm{Sr}}=0.10$ also
at low temperature, in the range 20~K~$\leq T \leq$~570~K. Sample
$x_{\rm{Sr}}=0.05$ was measured on warming from 30~K to 305~K with a
ramp of 0.8~K/min. Sample $x_{\rm{Sr}}=0.10$ was heated from 20~K to
305~K at 0.8~K/min, then up to 525~K at 1~K/min. The cooling ramp
was performed at 2~K/min down to 20~K and finally heated to RT at 0.8~K/min. In all cases the collected diffractograms
were saved every 2 min. Sample holders consisted on vanadium
cylinders that were not sealed, placed inside the furnace and
measured under a vacuum of $10^{-7}$ bar.

In addition, high--resolution data were collected at diffractometer
Super--D2B of ILL both before and after the experiment at D20,
\emph{i.e.}, after being heated to 570~K. A wavelength of $\sim
1.594$~{\AA} was used to obtain patterns at selected temperatures using the high--intensity configuration.
Details are described in \cite{07Aur2}. The neutron diffraction
patterns were processed with the full--pattern analysis Rietveld
method, using the program
\begin{scriptsize} FULLPROF
\end{scriptsize}~\cite{fullprofB} for refining the crystal and
magnetic structures. The strategy for the Rietveld refinement was as
follows. For the heating ramps, the starting point were the structures reported
previously~\cite{09Aur} obtained from high resolution D2B data, then used to refine sequentially the
neutron thermodiffractograms obtained at D20. Temperature scans
where divided into different ranges according to the structural and
magnetic order, and for each range the corresponding atomic
positions and occupations obtained at D2B were kept fixed (except
for oxygen at the adequate sites), while lattice parameters,
temperature factors and magnetic moments were allowed to vary. The
cooling ramps were refined based on the high resolution data from
D2B collected after the high--temperature experiments.

Magnetization measurements were performed using a
VSM and SQUID magnetometers at low temperature (5~K~$<T<$~310~K) and a
home-made Faraday balance at high temperature (300~K~$<T<$~600~K),
under a 5~kOe field, at a warming/cooling rate of 0.7~K/min in a
$1.33 \times 10^{-3}$ bar air atmosphere. Electrical resistivity
measurements were performed between 80~K and 300~K under
zero-applied magnetic field using the standard four-probe technique
with gold contacts sputtered onto the samples to reduce contact
resistance.

\section{Results \label{s:results}}

\subsection{Initial room temperature structures}

\begin{figure}[htb]
\centering
\includegraphics[width=\linewidth]{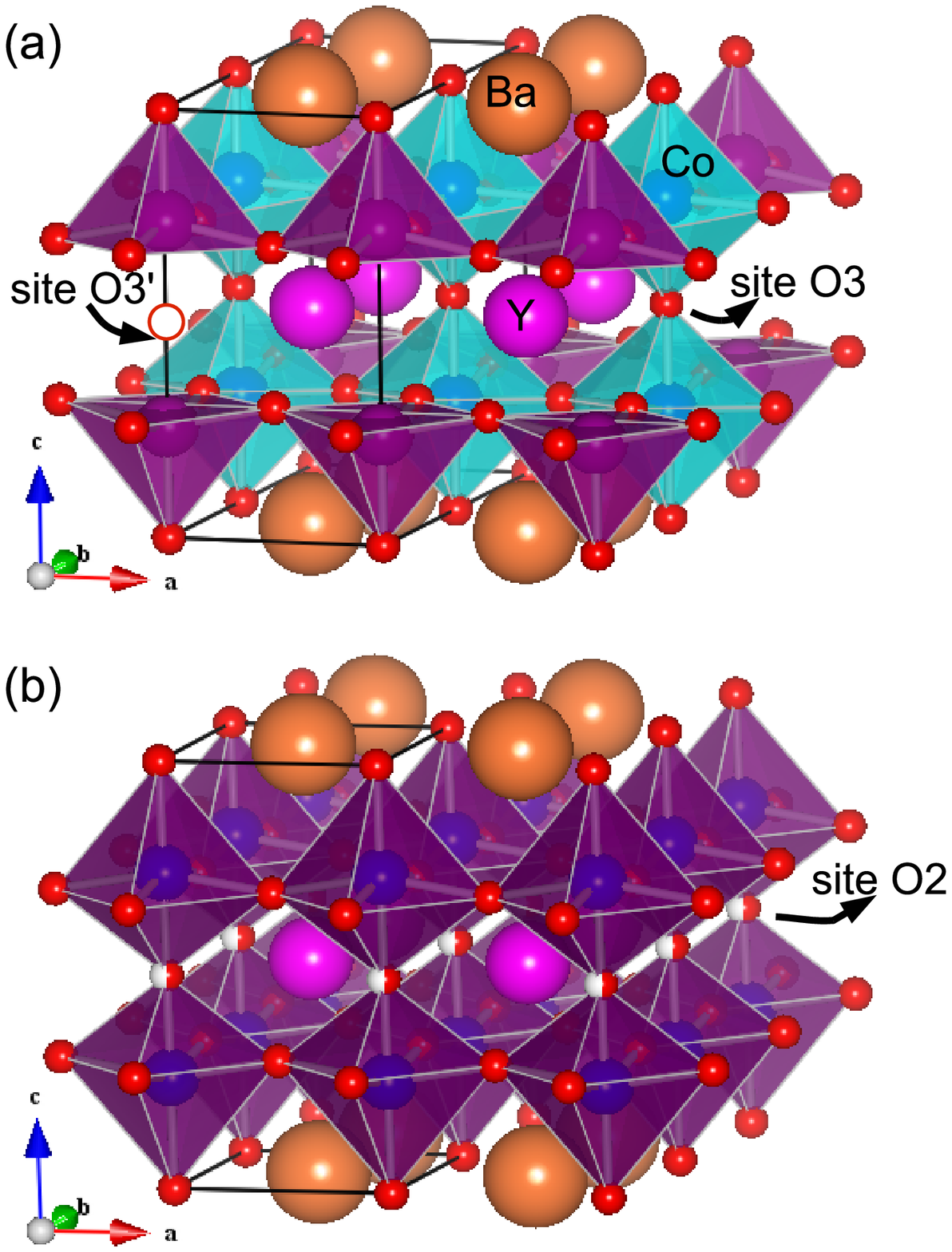}
\caption{(Schematic view of the structure for (a) the
O$_{122}$-phase with $\delta=0.5$ and complete order of oxygen vacancies
resulting in a larger cell and (b) the T$_{112}$-phase. The apical
oxygen partially colored in the graph (middle plane) indicates that
such site is partially occupied, leading to a random distribution of
oxygen vacancies. The crystallographic sites where oxygen vacancies distribute are also indicated, following the notation used in our previous articles \cite{07Aur2,09Aur}.} \label{f:1}
\end{figure}

In the first place, let us briefly summarize the starting point for
the three samples to be studied at high--temperature. The parent
compound, $x_{\rm{Sr}}=0$, has been so far intensively studied; and different structures as a function of temperature and oxygen content have been reported \cite{01Aka,05Pla,07Kha,08Kha,99Aka,09Koz,08Mal,10Pad}. Above RT it is in the orthorhombic $Pmmm$ space group with a ``122" cell (O$_{122}$-phase, Fig.~\ref{f:1}(a)). However, the substitution with Sr at the Ba site gradually favors the stabilization of a tetragonal $P4/mmm$
phase with ``112" structure, \emph{i.e.}, with no ordering of oxygen
vacancies~\cite{09Aur}. For sample $x_{\rm{Sr}}=0.05$ the RT state
corresponds to a mixture of 80 at.\% O$_{122}$-phase
and 20 at.\% tetragonal ``112"
phase (T$_{112}$-phase, Fig.~\ref{f:1}(b)). Sample
$x_{\rm{Sr}}=0.10$ is already completely tetragonal, showing no significant traces of O-phase neither in the neutron diffraction experiments nor in its magnetization as a function of temperature. In all cases, the initial oxygen
content refined from D2B data at and below RT is $\delta=0.47 \pm 0.02$.

\subsection{Evolution of phases upon heating \label{s:stab-w}}

\textit{Samples $x_{\rm{Sr}}=0$ and $x_{\rm{Sr}}=0.05$} \\

\begin{figure}[ptb]
\centering
\includegraphics[angle=-90,width=\linewidth]{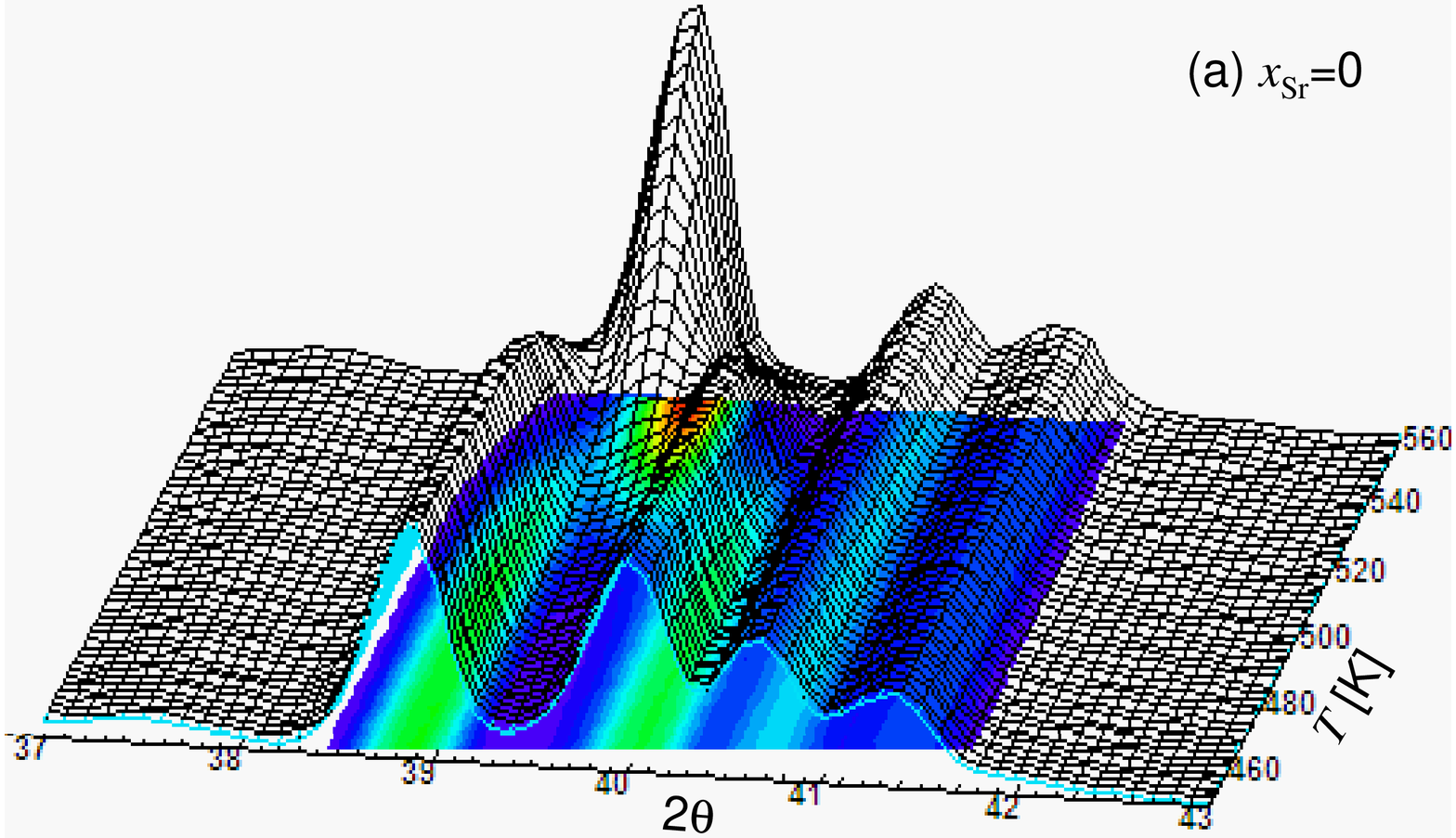}
\includegraphics[angle=-90,width=\linewidth]{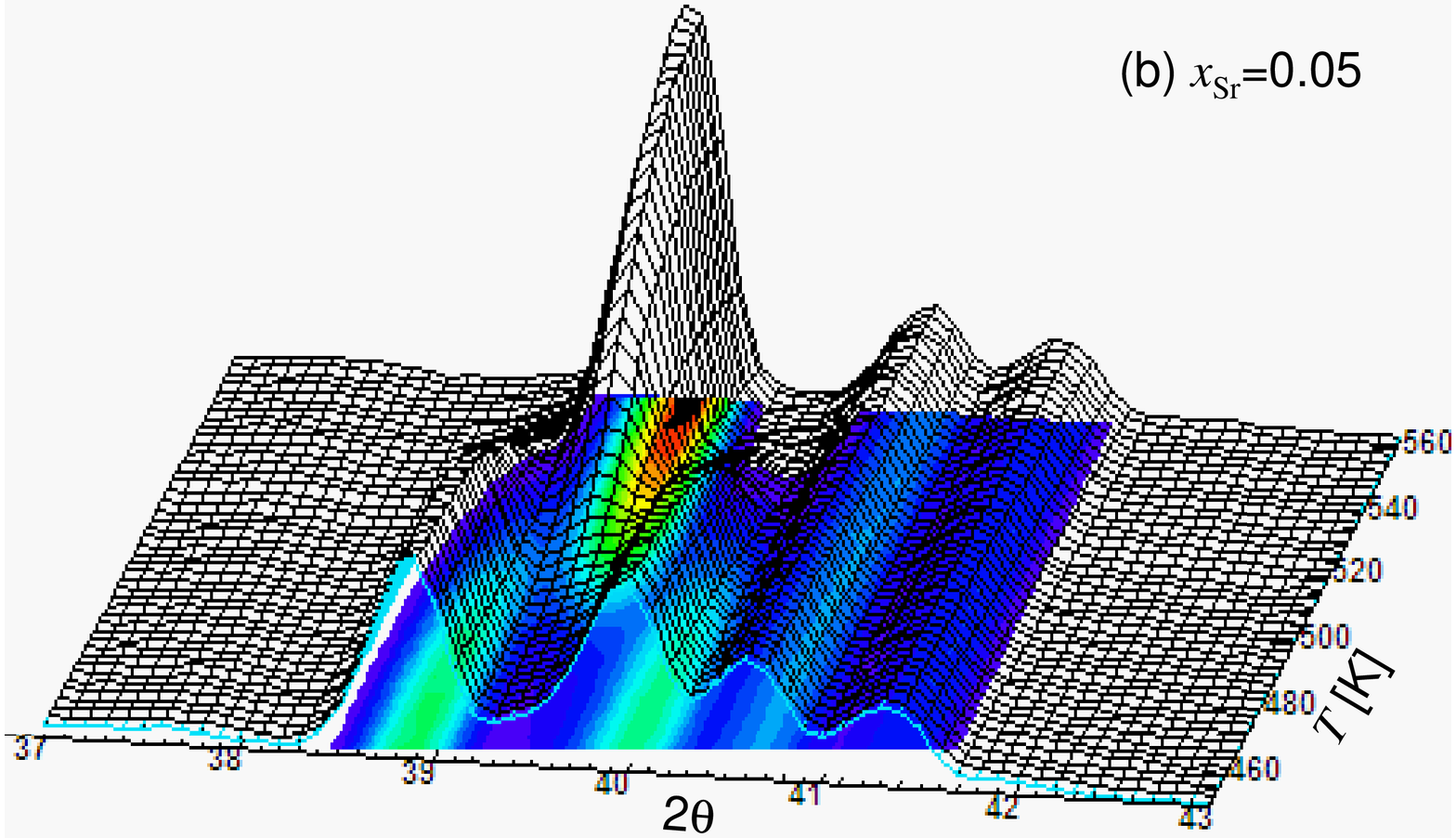}
\caption{ Thermodiffractograms and their projection to the $2 \theta
- T$ plane in samples with $x_{\rm{Sr}}=0$ (a) and
$x_{\rm{Sr}}=0.05$ (b), illustrating the orthorhombic to tetragonal transition. Data were collected on
warming at D20 with $\lambda \sim 1.31$ {\AA}. } \label{f:2}
\end{figure}

In Fig.~\ref{f:2}(a) we present the thermodiffractogram of sample
$x_{\rm{Sr}}=0$ collected on heating from RT to 570~K together with
the corresponding projection to the $2\theta - T$ plane, for a
selected region showing the Bragg reflections (0~4~0)$_{\rm{O}}$ and
(2~0~0)$_{\rm{O}}$ collapsing into the (2~0~0)$_{\rm{T}}$
reflection. This corresponds to a structural phase transition from the $Pmmm$ to the $P4/mmm$ space group ocurring at $\backsim$
530~K . A similar
orthorhombic to tetragonal high-temperature transition has been reported in this
family of compounds by Streule \emph{et al.}~\cite{06Str} for
\emph{R}=Pr, but occurring at a higher temperature, \emph{viz.},
776~K. That transition corresponds to a disordering of
vacancies, given that their measurements were performed in sealed sample
holders filled with an inert gas preventing any oxygen loss
during the experiment. As we will show below, in our experiment the
transition is driven by the oxygen loss and rearrangement
of oxygen vacancies.

\begin{figure}[ptb]
\centering
\includegraphics[angle=-90,width=\linewidth]{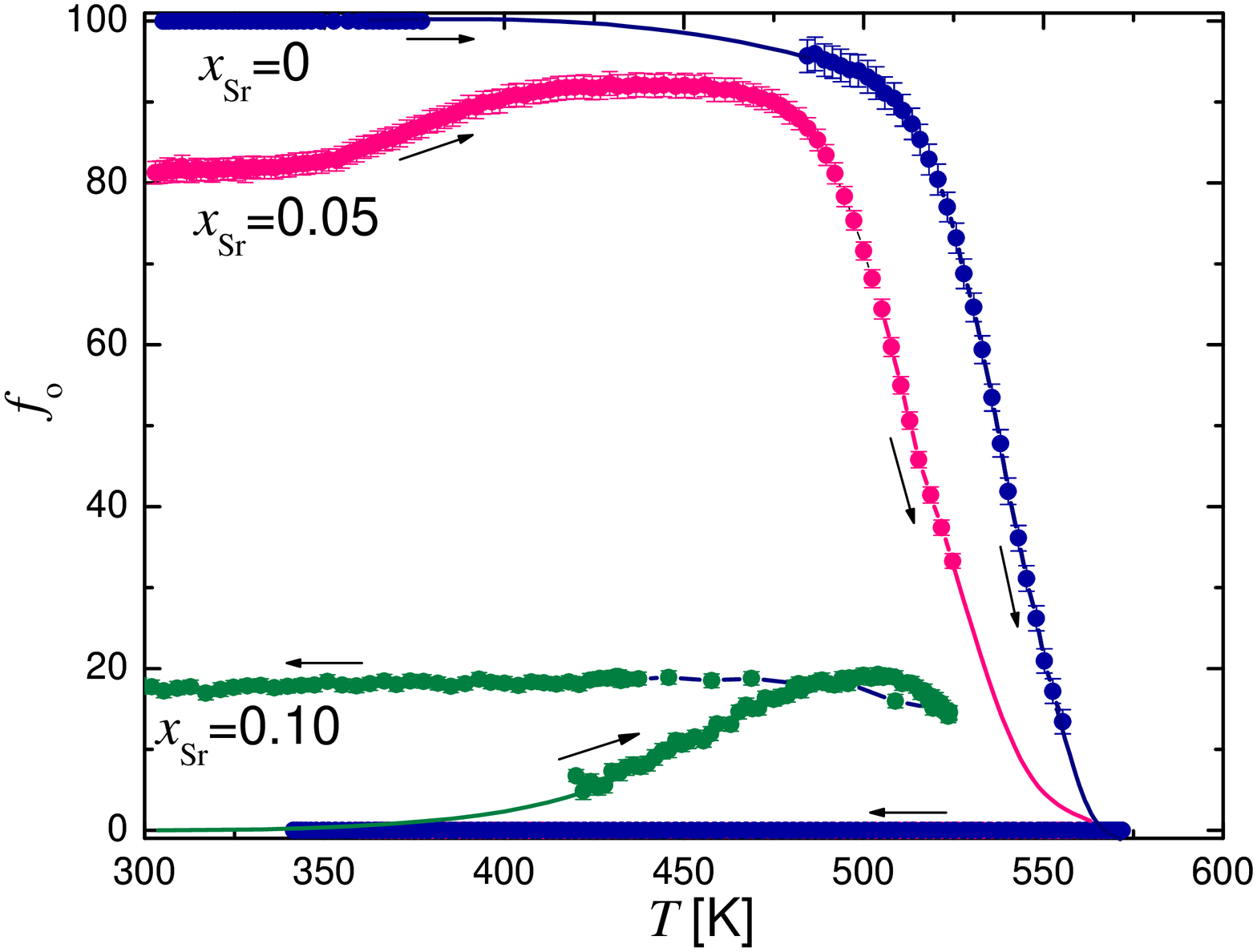}
\caption{ Phase fraction of the O$_{122}$-phase
above RT for samples $x_{\rm{Sr}}=0$, $x_{\rm{Sr}}=0.05$ and
$x_{\rm{Sr}}=0.10$. Arrows indicate if data were collected on
warming or on cooling.} \label{f:fract}
\end{figure}

The sample with $x_{\rm{Sr}}=0.05$ shows a similar evolution with
temperature, as shown in Fig.~\ref{f:2}(b). The
transition here occurs at a lower temperature ($\sim 510$~K), which is
consistent with the fact that Sr addition already helps to stabilize the
tetragonal structure. However, we observe a slight but noticeable
increase in the O$_{122}$ phase fraction with increasing temperature
between 350~K~$\leq T \leq$~400~K, which can be interpreted as the
gradual stabilization of the O$_{122}$ phase due to oxygen mobility
and its tendency to become ordered, until the oxygen loss leads to
the complete transformation to the vacancy-disordered T$_{112}$
phase. The evolution of the phase fractions for samples
$x_{\rm{Sr}}=0$ and $x_{\rm{Sr}}=0.05$ is presented in
Fig.~\ref{f:fract}. At the end of the temperature ramp, both samples were
kept at 570~K for 30 minutes and then cooled down to RT with a
controlled ramp. The transformation to the
T-phase was already complete and irreversible. This
was expected considering that the furnace was sealed and with a
vacuum of $10^{-7}$ bar so that oxygen was irreversibly lost.

\begin{figure}[ptb]
\centering
\includegraphics[width=\linewidth]{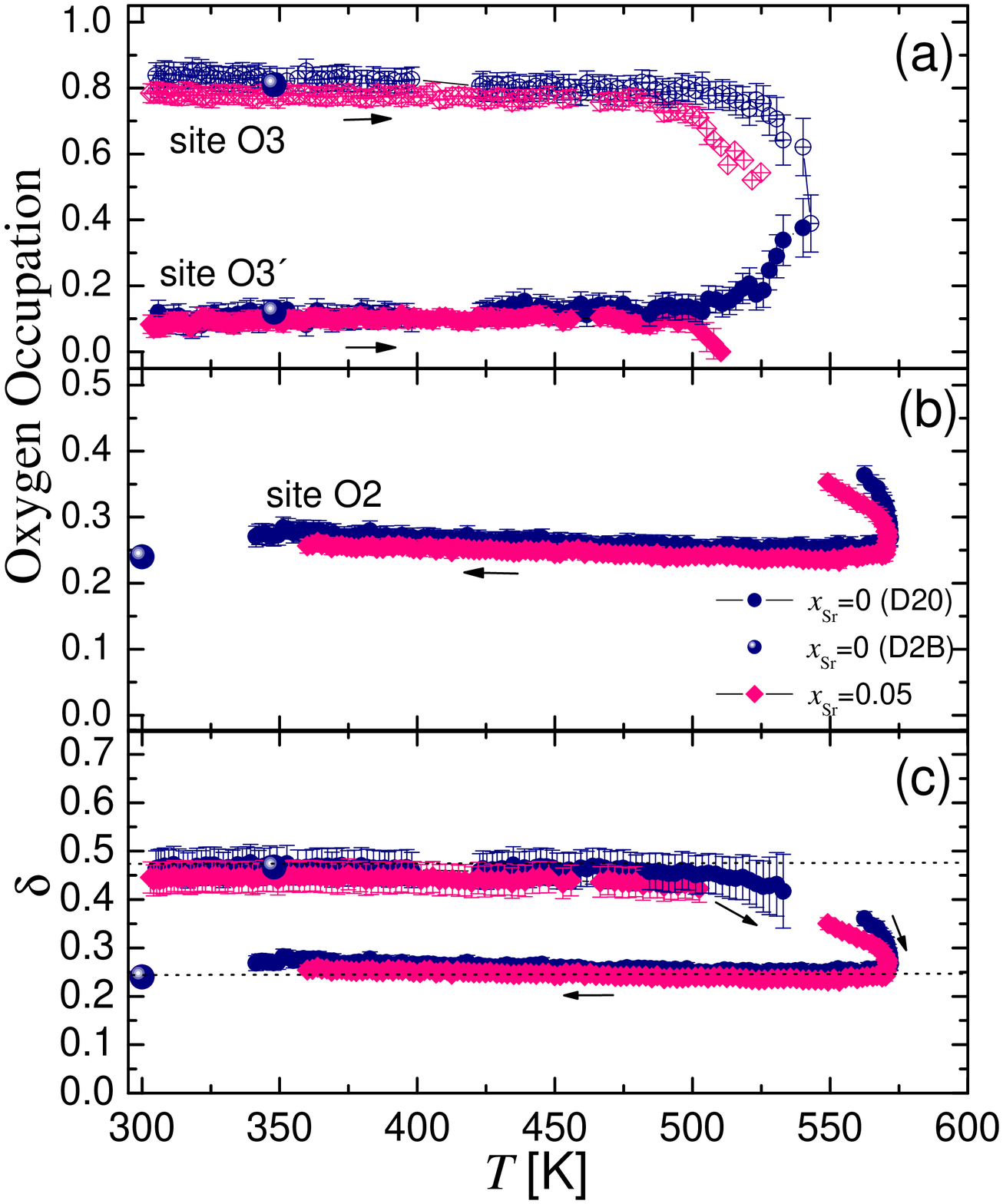}
\caption{(a) Oxygen occupancies obtained from the refinements for sites O3 and O3' in the [YO$_{\delta}$] plane of the
O$_{122}$ phase for samples $x_{\rm{Sr}}=0$ and $0.05$. (b) Oxygen
occupancy of the O2 site in the [YO$_{\delta}$] plane of the
T$_{112}$ phase. (c) Global oxygen content in samples
$x_{\rm{Sr}}=0$ and $0.05$ for the whole warming-cooling process.}
\label{f:occ}
\end{figure}

To study the evolution of oxygen vacancies during the heating-dwelling-cooling process, we analysed the occupancies of the relevant crystallographic sites occupied by oxygen atoms, obtained from the
refinement of D20 data. In the notation adopted for the present work, the sites O3$/$O3' of the $Pmmm$ space group (O$_{122}$ phase) are occupied/unoccupied by oxygen atoms located in the [YO$_{\delta}$] plane. They correspond to the apical oxygens which define octahedra and pyramids (Whykoff positions $1g$ and $1c$ respectively) and they are indicated in Fig.~\ref{f:1}(a). In the ideal ``122" structure the site O3 is completely occupied (50\% octahedra) and site O3' is completely empty (50\% pyramids), although it is frequent to find a small degree of disorder even at low temperature~\cite{07Aur2}. In the present refinements, we have allowed for disorder in this array by setting these occupancies as free parameters. The results are presented in Fig.~\ref{f:occ}(a) for the $x_{\rm{Sr}}=0$ and $0.05$ samples. The elevated degree of order among pyramids and octahedra remains unaltered just until the transition at 530~K. At this temperature, we observe for the sample $x_{\rm{Sr}}=0$ a sudden rearrangement of vacancies, leading to both sites having the same occupancy (0.5), in agreement with the transition to the T$_{112}$ phase in which pyramids and octahedra are distributed randomly. For the sample $x_{\rm{Sr}}=0.05$ the same rearrangement occurs, although it is not so neatly observed: it should be noticed that in this sample there is already a fraction of T phase in coexistence blurring the results. \\
After the transition, data were refined in the tetragonal $P4/mmm$ space group. In this description, oxygen vacancies are still located in the [YO$_{\delta}$] plane defining the apices of the octahedra, but as there is no particular order there is only one crystallographic site for apical oxygen in this structure. In the present notation it is called O2 site (Whykoff position $1b$). When the O2 site is completely occupied then $\delta=1$, when completely unoccupied $\delta=0$. Figure~\ref{f:occ}(b) shows the evolution of the O2 site occupancy in the T-phase after the transition. It can be observed that the samples loose oxygen during the final steps of the heating ramp, decreasing the occupancy of the O2 site, and they continue to do so during the 30 minute dwell at 570~K. The bottom
panel (Fig.~\ref{f:occ}(c)) corresponds to the evolution of the $\delta$ value for both samples. It has been calculated combining data on site occupancies (Occ) and phase fractions ($f$), following the equation
\begin{equation*}
 \delta=f_{\mathrm{O}} \times \frac{\left( \mathrm{Occ}_{\mathrm{O3}}+ \mathrm{Occ}_{\mathrm{O3'}}\right)} {2} + f_{\mathrm{T}} \times  \mathrm{Occ}_{\mathrm{O2}}.
\end{equation*}

\begin{figure}[ptb]
\centering
\includegraphics[width=0.8\linewidth,angle=-90]{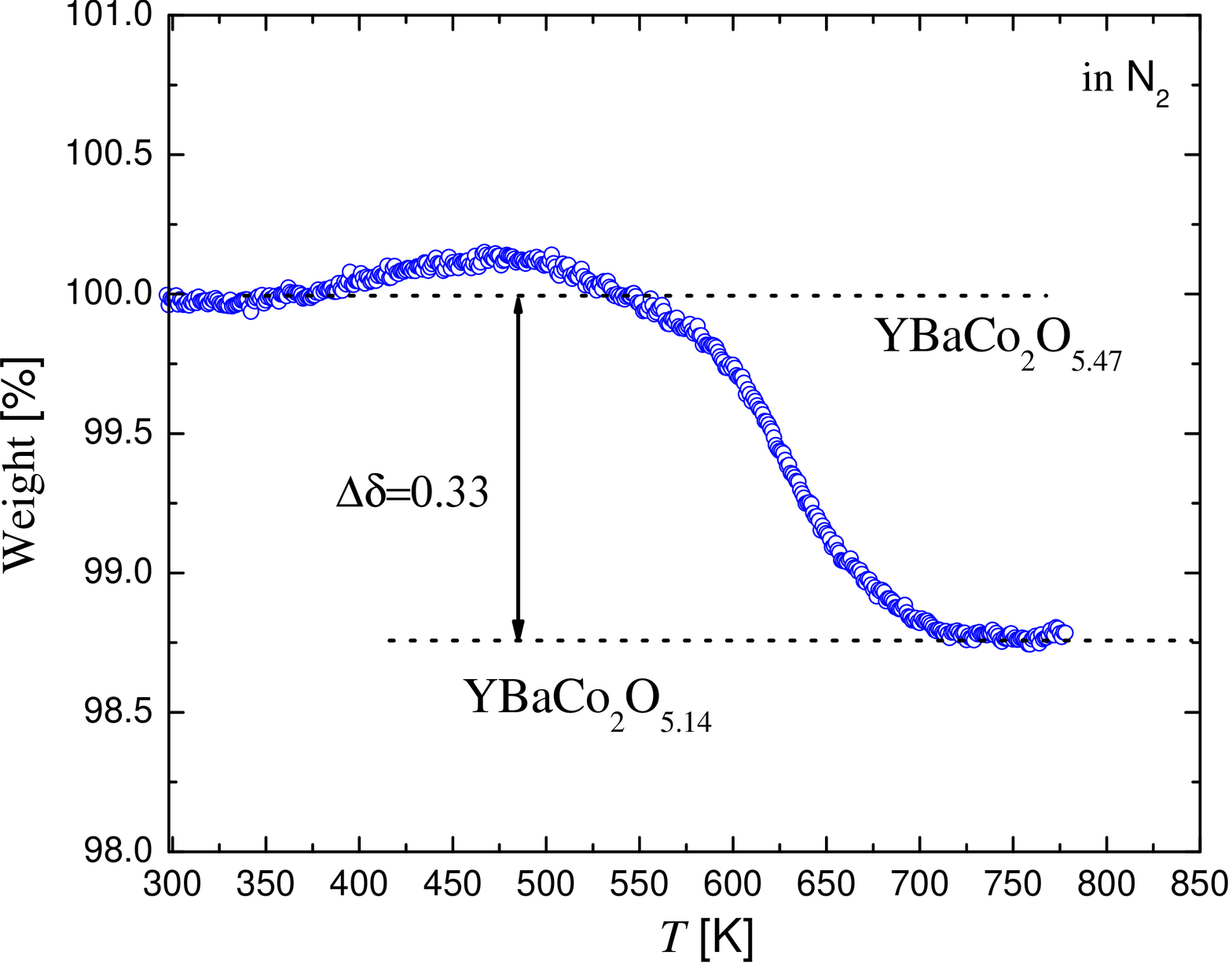}
\caption{Thermogravimetric curve for a sample of YBaCo$_2$O$_{5.47}$ performed under nitrogen flow. The initial and final states are indicated.}
\label{f:TG}
\end{figure}

We observe in Fig.~\ref{f:occ}(c) a smooth transition from the initial value 0.47 to a final value of 0.24 in both samples.
The same final value for the oxygen content was also obtained from the refinement of RT high-resolution data collected after the D20 experiments, even after a further structural transition occurs during the cooling ramp, as we will show in the following section, giving consistency to the results. Certainly, a different oxygen content would be atteined by varying the dwell time, until an equilibrium is reached at 570~K under the conditions of vacuum used in the experiment. A thermogravimetric analysis was performed on the parent compound, under a nitrogen flow of
60 cm$^{3}/$min and with a heating rate of $2^{\circ}/$min from 298~K to
773~K. Figure~\ref{f:TG} shows that the oxygen loss under nitrogen flow begins at around 550-600 K, in good agreement with Ref.~\cite{99Aka}. In such conditions, the sample ends with a value of $\delta$ close to $0.14$ under the assumption that the starting oxygen content corresponds to that obtained from the neutron diffraction experiments and that all the weight lost corresponds to oxygen atoms. Note that this final oxygen content is stable from $\approx 700$~K. Moreover, it is expected that the cobaltites present a larger oxygen loss under nitrogen than under oxygen flow~\cite{07Hao}, so the final $\delta$ value in both exeperiments seems consistent. For the objectives of the present work, the irreversible oxygen loss in our neutron diffraction experiment was exploited to compare the parent compound and the 5\% Sr-substituted samples for an oxygen content much lower than $\delta=0.5$. In addition, the thermal treatment mimics the one performed during the experiments at the Faraday balance (described in the following) and help us understand the behavior observed after cooling again to RT.

\textit{Sample $x_{\rm{Sr}}=0.10$}

\begin{figure}[ptb]
\centering
\includegraphics[angle=-90,width=\linewidth]{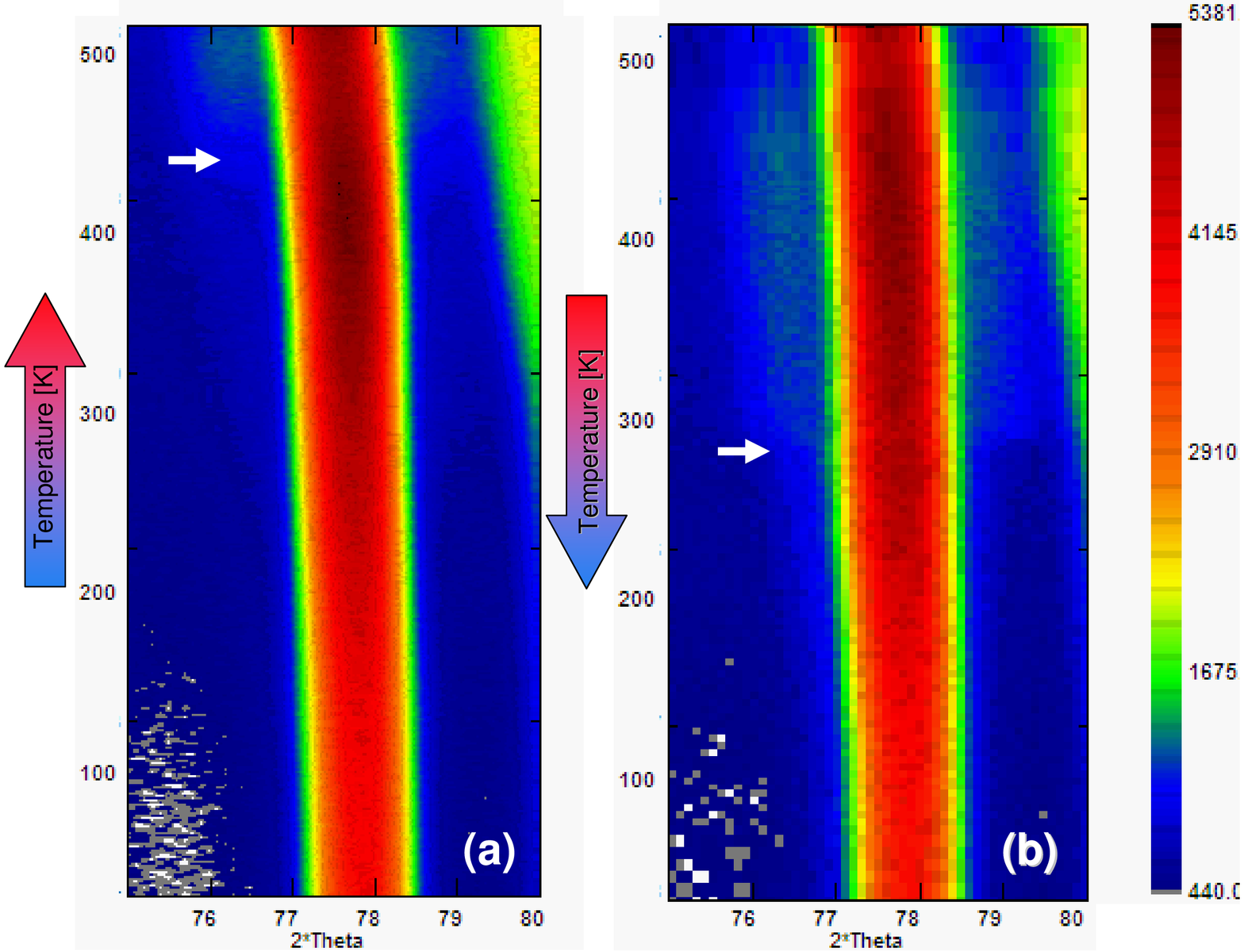}
\caption{Projection to the $2 \theta - T$ plane of the
thermodiffractograms for sample $x_{\rm{Sr}}=0.10$ showing the major
(2~0~0) Bragg reflection of the T$_{112}$-phase and the onset of the
(2~0~0) and (0~4~0) Bragg reflections of the O$_{122}$ structure.
Data were collected on warming (a) and cooling (b) at D20 with
$\lambda \sim 2.42$ {\AA} between 300~K and 525~K.}
\label{f:ramp-VIII}
\end{figure}

As we mentioned above, the initial state for this sample corresponds completely to the T$_{112}$-phase
although the initial oxygen content is the same as in the other samples. This means that pyramids
and octahedra in the structure are still in a ratio close to $1:1$
but, due to the disorder introduced by Sr replacement, they are not anymore ordered along the $b$-axis. However, we observed at high--temperature the unexpected formation of the ordered O$_{122}$-phase around 425~K. To illustrate this, the diffraction data collected on warming projected to the $2\theta - T$ plane, for the Bragg reflection (2 0
0)$_{\rm{T}}$ are presented in Fig.~\ref{f:ramp-VIII}(a). The arrow indicates the onset of the O$_{122}$ phase, showing up as rather diffuse shadows (actually, the (0 4 0)$_{\rm{O}}$
and (2 0 0)$_{\rm{O}}$ Bragg reflections of the O$_{122}$-phase) at each side of the major (2 0 0)$_{\rm{T}}$ reflection. As in the former
samples, this small fraction of O-phase begins to transform to a T-phase at $\backsim$ 500~K  but the transformation could not proceed because the maximum temperature reached by the furnace this time was 525~K for technical reasons. This can be observed in Fig.~\ref{f:fract} for the phase fraction of the O-phase. The warming ramp ends with a sample which is 85\% in the T-phase and
$~15$\% in the O-phase. But most importantly, there occurs no oxygen loss because the heating stopped before reaching the
onset of the desorption, and we immediately started the
cooling ramp without a dwell. In Fig.~\ref{f:ramp-VIII}(b), where the data collected on cooling are presented, we can see how this phase is
retained until the typical distortion of this phase occurs near RT (at the metal-insulator transition temperature) indicated by
the arrow in (b) and discussed in the next section.

\subsection{Structures retained on cooling \label{s:stab-c}}

\textit{Samples $x_{\rm{Sr}}=0$ and $x_{\rm{Sr}}=0.05$ with $\delta=0.24$}

\begin{figure}[ptb]
\centering
\includegraphics[angle=-90,width=\linewidth]{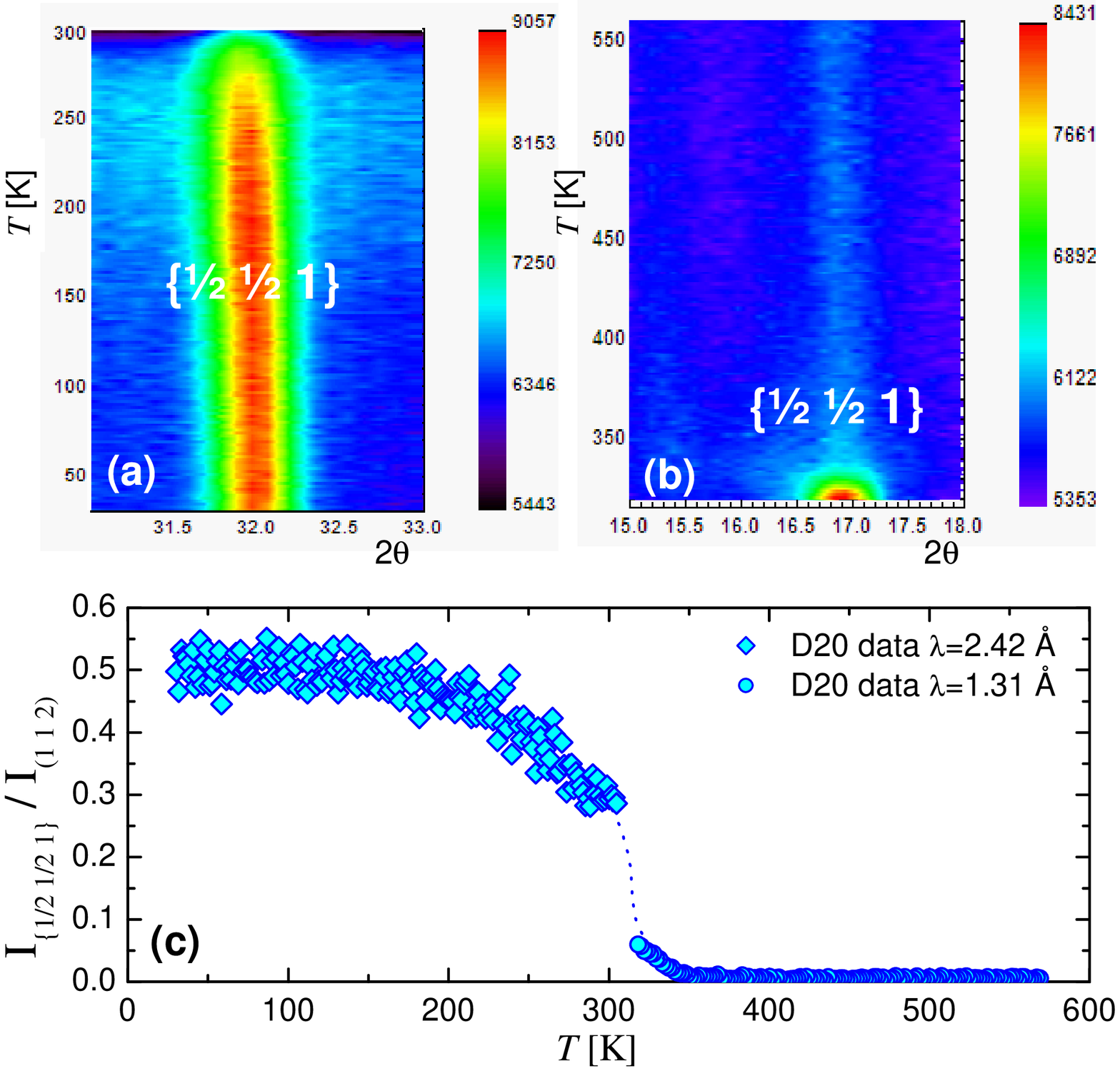}
\caption{(a,b) Projection of a selected section of the
thermodiffractograms of sample $x_{\rm{Sr}}=0.05$ showing the major
magnetic family of reflections $\{\frac{1}{2} \frac{1}{2}
1\}_{\rm{AFM-G}}$. Data in (a) were collected at D20 with $\lambda
\sim 2.42$ {\AA} in the temperature range 30~K to 300~K after the
high--temperature experiment. Data in (b) were collected during the
high--temperature experiment, on cooling, with a $\lambda \sim 1.31$
{\AA}. (c) Thermal evolution of the intensity of the most intense
magnetic peak ($\{\frac{1}{2} \frac{1}{2} 1\}_{\rm{AFM-G}}$)
relative to the most intense nuclear reflection (1 1 2)$_{\rm{T}}$,
in the sample $x_{\rm{Sr}}=0.05$. } \label{f:mag}
\end{figure}

\begin{figure}[ptb]
\centering
\includegraphics[angle=-90,width=\linewidth]{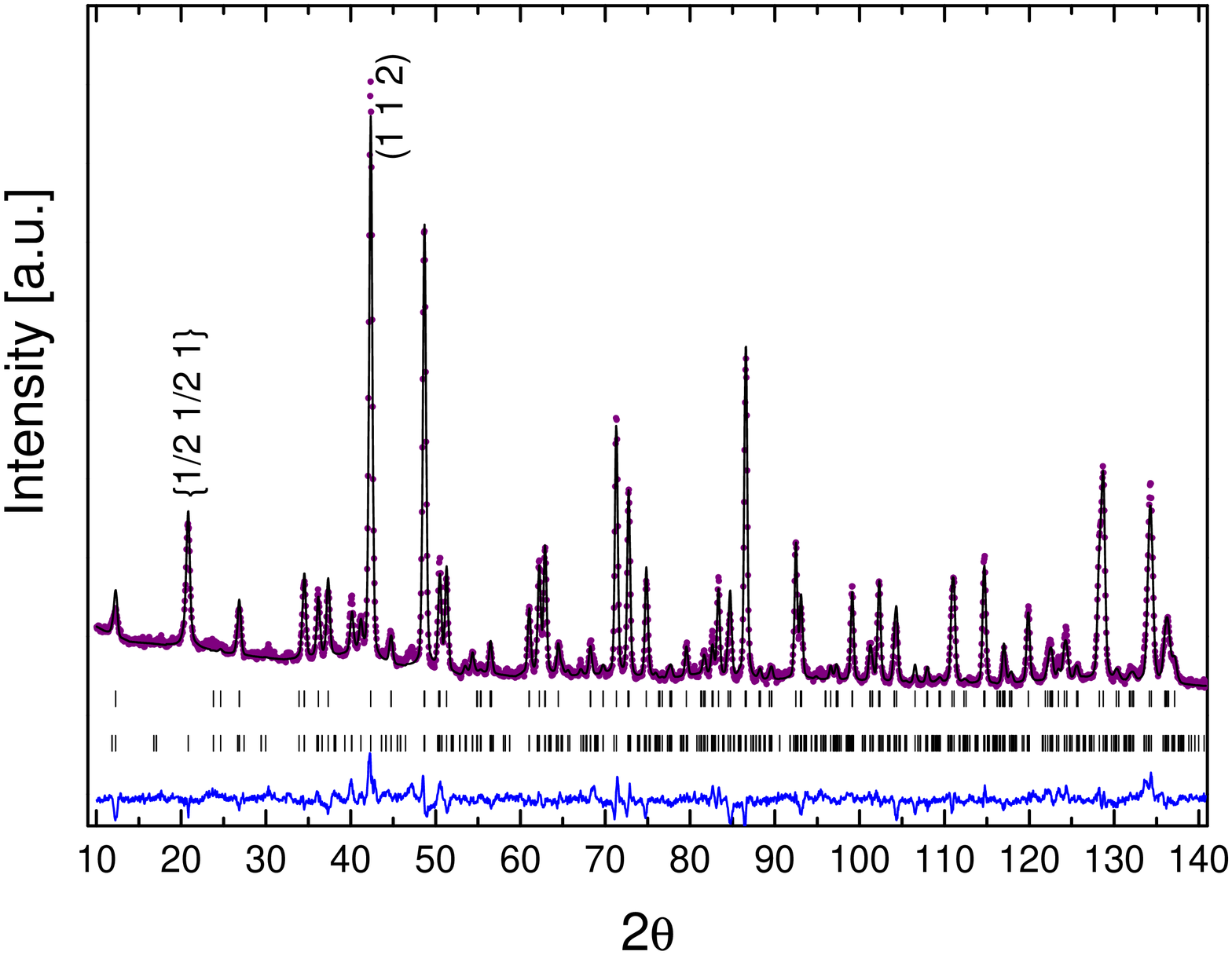}
\caption{ Rietveld refinement for the sample with $x_{\rm{Sr}}=0.05$ and $\delta=0.24$
from data collected at D2B at $T=100$~K. Vertical bars at
the bottom indicate Bragg reflections from the phases included in the
refinement: the nuclear phase $Pmmm$ ``112" and the G-type AFM phase ``222".} \label{f:refinement}
\end{figure}

Samples $x_{\rm{Sr}}=0$ and $x_{\rm{Sr}}=0.05$ were kept for 30
minutes at 570~K and then cooled at the same rate (1.2~K/min) down
to 343~K ($x_{\rm{Sr}}=0$) and 313~K ($x_{\rm{Sr}}=0.05$). No
transformations were observed upon cooling for the $x_{\rm{Sr}}=0$
sample. On the other hand, the onset of a magnetic ordering was
observed in the $x_{\rm{Sr}}=0.05$ sample which was cooled to a
lower temperature. This magnetic transition above RT corresponds to a G-type AFM ordering as observed in YBaCo$_2$O$_5$
\cite{00Vog}. It is illustrated in Fig.~\ref{f:mag} where we present
in panel (b) a projection of the family of magnetic
reflections $\{\frac{1}{2} \frac{1}{2} 1\}$. In panel (c)
the intensity of this magnetic peak is plotted normalized to the
most intense nuclear reflection (1 1 2). The data presented below RT
and the projection in Fig.~\ref{f:mag}(a) correspond to the second
experiment performed in D20 but with a wavelength of 2.42~\AA. We
can see a smooth transition between both data sets and interpolate
them to estimate the magnetic transition temperature. The transition
also occurs in our parent compound sample, but unfortunately we
could not reach the transition temperature on cooling. However, the
high-resolution diffractogram collected at D2B after the experiment
confirms that there is a G-type AFM order in the structure.
It has been reported by Vogt \emph{et al.}~\cite{00Vog} that the
magnetic transition is accompanied by a tetragonal to orthorhombic
distortion. We have therefore refined the high-resolution data
for samples $x_{\rm{Sr}}=0$ and $x_{\rm{Sr}}=0.05$ in the
orthorhombic $Pmmm$ space group (of course as $\delta=0.24$ the cell
is now ``112") and using a G-type AFM magnetic model for $T=100$~K, 250~K and
300~K. In Fig.~\ref{f:refinement} we show one of such refinements for sample $x_{\rm{Sr}}=0.05$ at 100~K. The most intense magnetic and nuclear reflections are indicated. In Table~\ref{t:1} the structural details
of the $Pmmm$ phase in our samples (for $\delta=0.24$) are
presented, obtained from the refinement of high-resolution data.

\begin{table}[bt]
\caption{Structural parameters of the O$_{112}$-phase refined from
the high resolution D2B data for the compound YBaCo$_2$O$_{5.25}$ at
$T=300$~K and YBa$_{0.95}$Sr$_{0.05}$Co$_2$O$_{5.25}$ at $T=100$~K,
250~K and 300~K. Atomic fractional coordinates correspond to space
group $Pmmm$ in the following Wyckoff positions: Y
(\emph{1h})=($\frac{1}{2},\frac{1}{2},\frac{1}{2}$); Ba, Sr
(\emph{1f})=($\frac{1}{2}, \frac{1}{2}, 0)$; Co (\emph{2q})=
($0,0,z$); O1 (\emph{1a})=($0,0,0$); O2
(\emph{2s})=($\frac{1}{2},0,z$); O3 (\emph{2r})=($0,\frac{1}{2},z$);
O4 (\emph{1c})=($0,0,\frac{1}{2}$).} \label{t:1}
\begin{indented}
\lineup
\item[]
\begin{tabular}{@{}lcccccc}
\br & & $x_{\rm{Sr}}=0$ & &  & $x_{\rm{Sr}}=0.05$ & \\
 & & $T=300$~K &  & $T=300$~K & $T=250$~K & $T=100$~K\\ \mr
Co & $z$ & 0.259(1)  & & 0.2584(6)  & 0.2584(6) & 0.2572(6) \\
O2 & $z$ & 0.3111(8)  & & 0.3116(7)  & 0.3110(5) & 0.3101(5) \\
O3 & $z$ & 0.2984(8)  & & 0.2963(4)  & 0.2979(5) & 0.2990(5) \\
O4 & $Occ$ & 0.24(1)  & & 0.24(1)  & 0.24(1) & 0.23(1) \\
$a$ ({\AA}) &  & 3.8752(6) & & 3.8731(2)  & 3.8745(2) & 3.8720(2) \\
$b$ ({\AA}) &  & 3.8777(6) & & 3.8763(2)  & 3.8717(2) & 3.8661(2) \\
$c$ ({\AA}) &  & 7.4992(4) & & 7.4993(2)  & 7.4910(4) & 7.4754(4) \\
$\mu_x$ ({\AA}) &  & 0.70(5) & & 0.76(5)  & 0.96(5) & 1.18(5) \\
$R_{\rm{B}}$ &   & 10.2 &  & 5.6  & 8.5 & 9.4 \\
$R_{\rm{mag}}$ &  & 18  & & 8.5  & 17 & 14 \\
$\chi^2$ &  & 1.4  & & 2.9  & 3.1 & 5.6 \\
\mr Co-O distances  ({\AA}) & &  & &  & & \\
$d_{\rm{Co-O1}}$ (ap) & & 1.945(7) & & 1.938(4)& 1.936(4) & 1.923(4) \\
$d_{\rm{Co-O2}}$ (pl) & & 1.976(2) & & 1.977(1)& 1.977(1) & 1.976(1) \\
$d_{\rm{Co-O3}}$ (pl) & & 1.961(2) & & 1.959(1)& 1.958(1) & 1.957(1) \\
$d_{\rm{Co-O4}}$ (ap) & & 1.804(7) & & 1.812(4)& 1.810(4) & 1.815(4) \\
\br
\end{tabular}
\end{indented}
\end{table}

\textit{Sample $x_{\rm{Sr}}=0.10$ and $\delta=0.47$}

\begin{figure}[ptb]
\centering
\includegraphics[angle=-90,width=\linewidth]{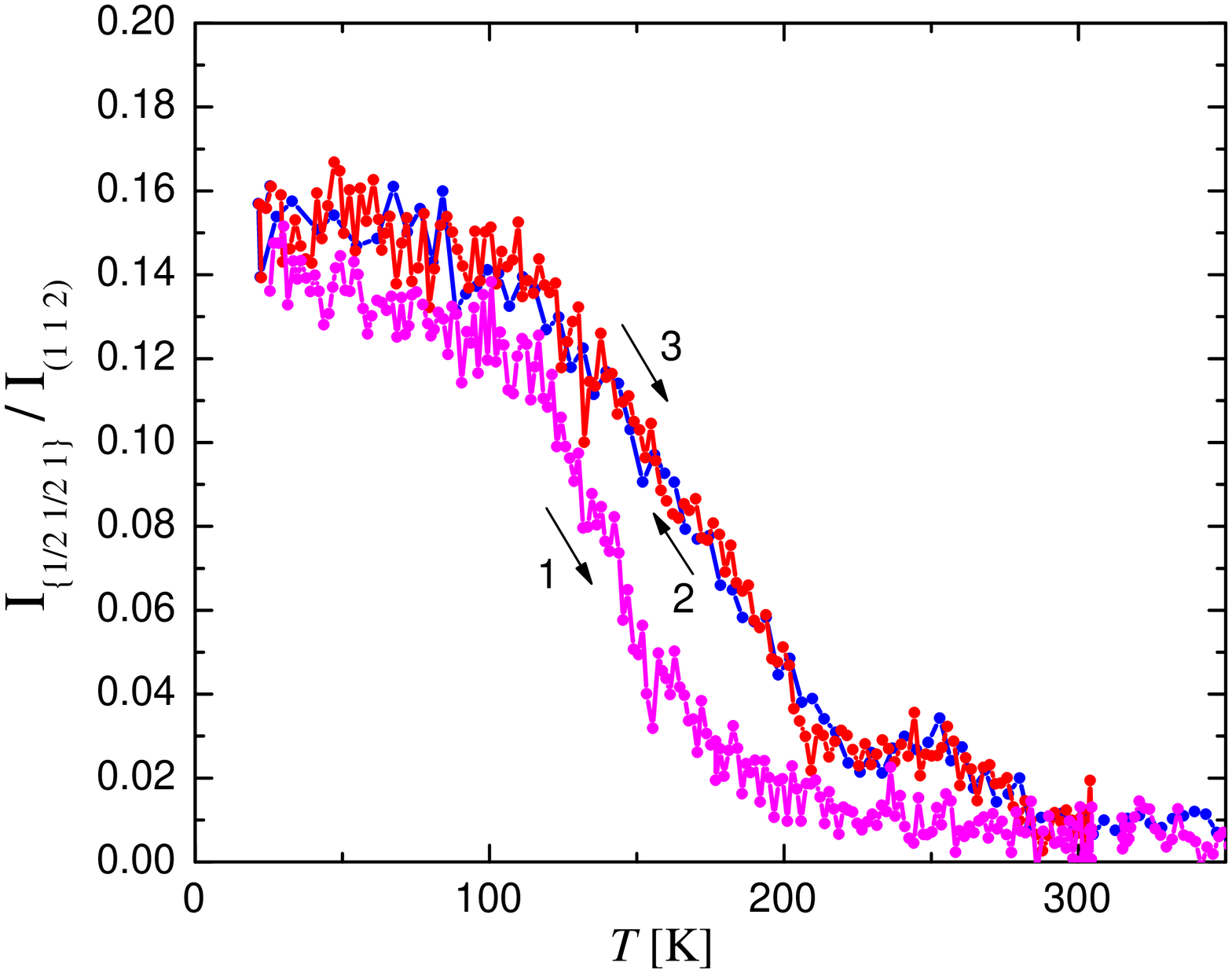}
\caption{Thermal evolution of the intensity of the magnetic family
of reflections $\{111\}_{\rm{ferri}}$ and/or $\{\frac{1}{2}
\frac{1}{2} 1\}_{\rm{AFM-G}}$, normalized to the most intense
nuclear reflection for sample $x_{\rm{Sr}}=0.10$ below RT. Numbered
arrows indicate the sequence of treatments applied to the sample as
explained in the text.} \label{f:mag-int}
\end{figure}

\begin{figure}[ptb]
\centering
\includegraphics[angle=-90,width=0.6\linewidth]{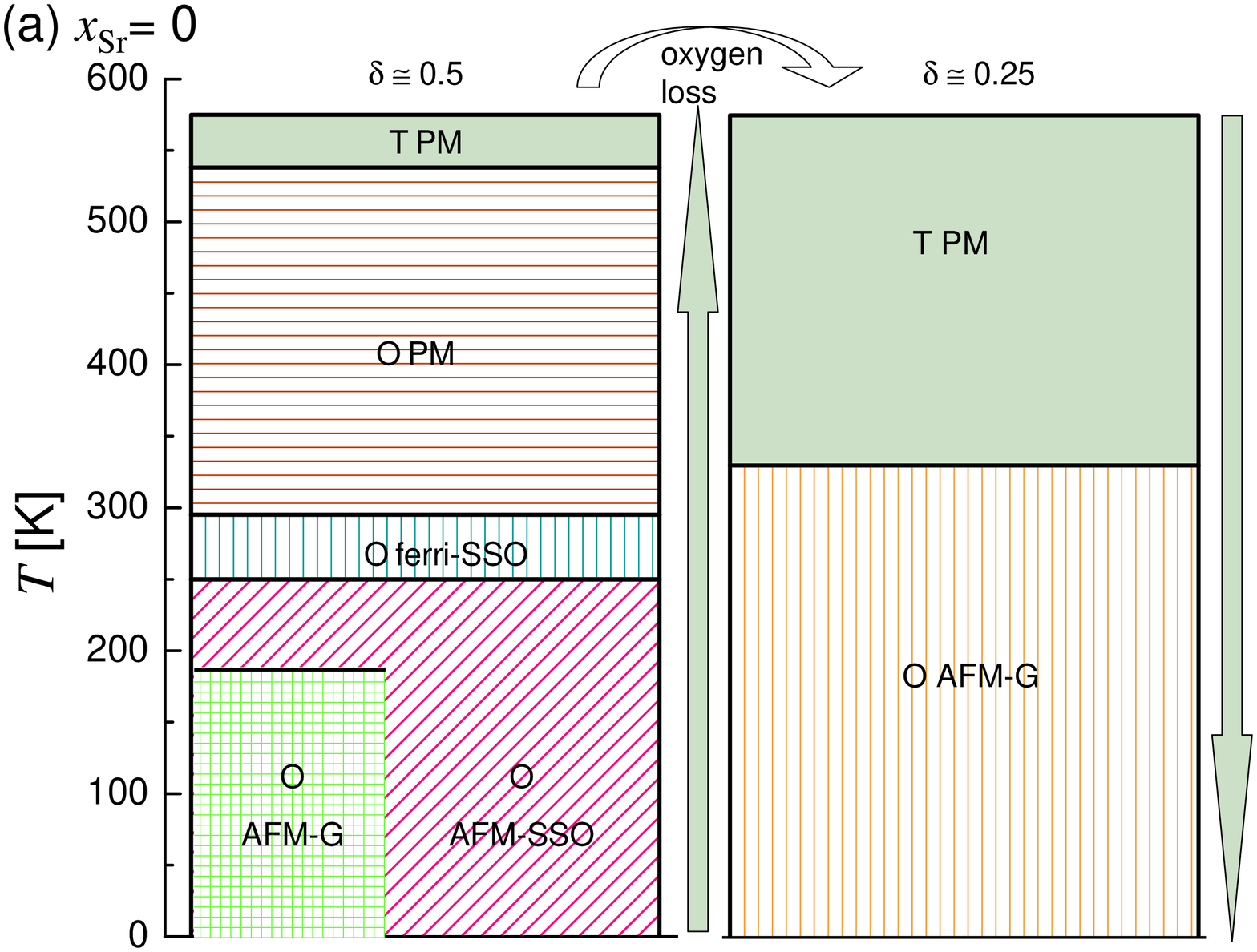}
\includegraphics[angle=-90,width=0.6\linewidth]{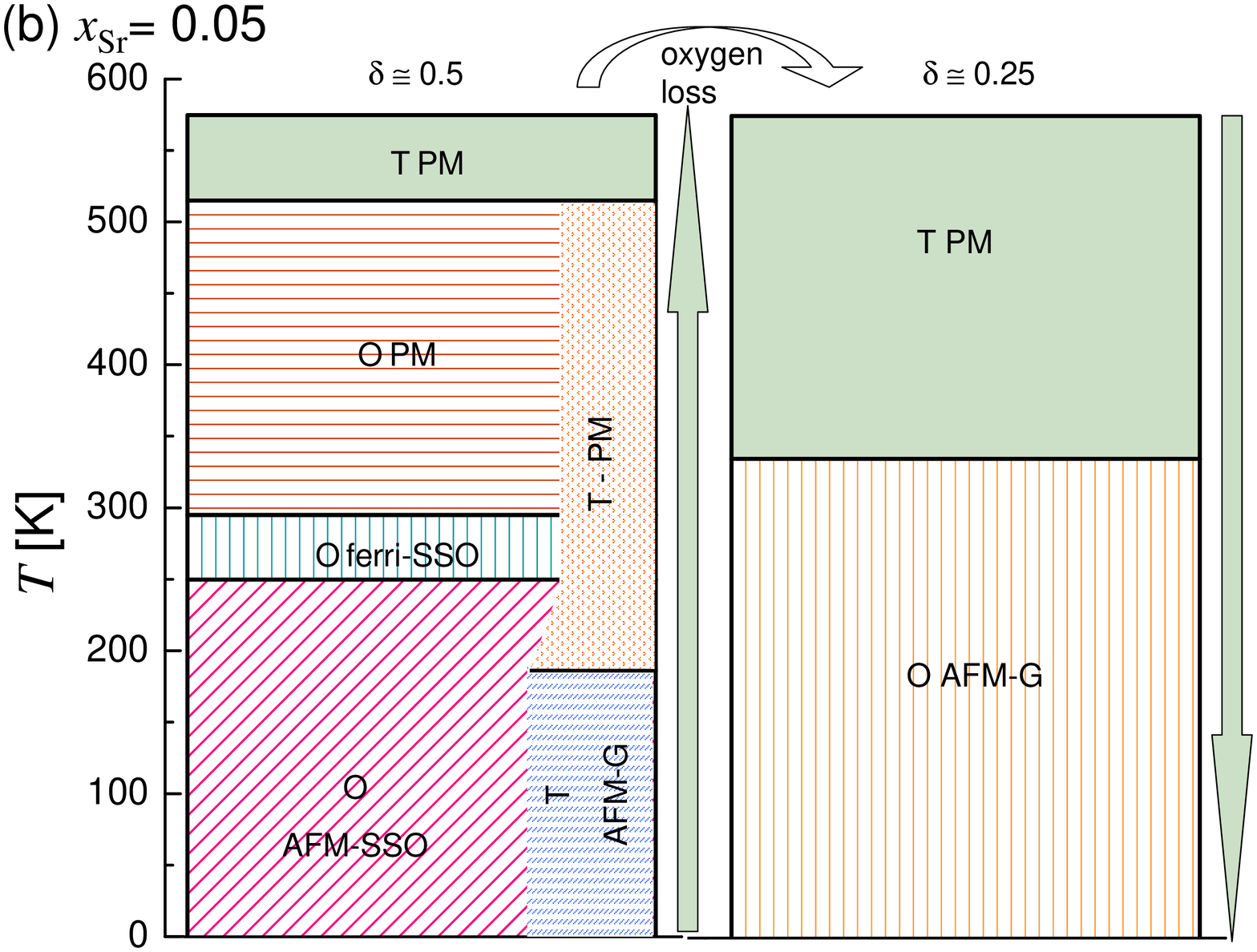}
\includegraphics[angle=-90,width=0.6\linewidth]{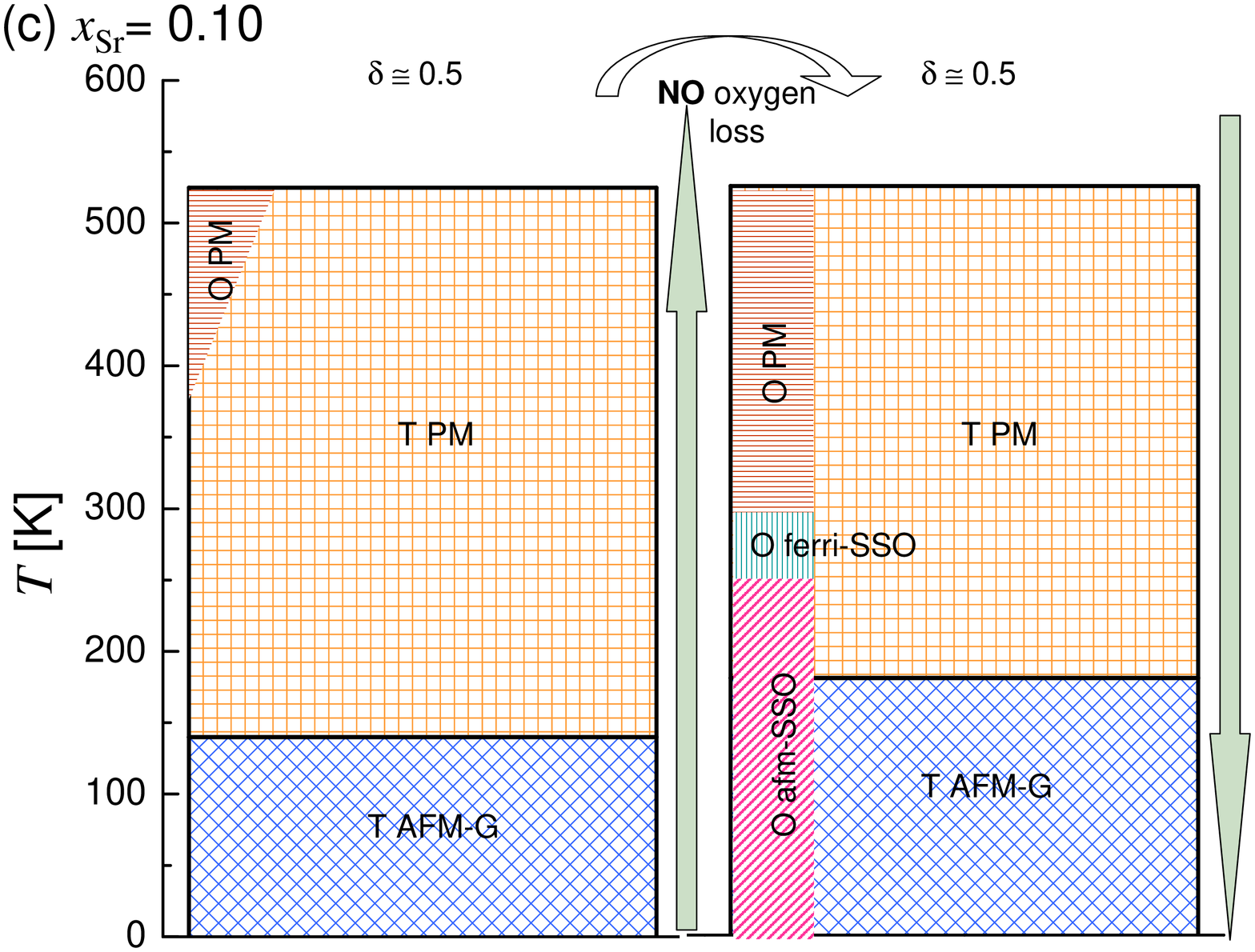}
\caption{Schematic diagram of the observed phases in the current experiment, for samples with $x_{\rm{Sr}}=0$
(a), $x_{\rm{Sr}}=0.05$ (b) and $x_{\rm{Sr}}=0.10$ (c). The column
on the left corresponds to the sequence of nuclear and magnetic
phases observed on warming, while the column to the right indicates
the sequence on cooling. For samples $x_{\rm{Sr}}=0$ (a) and
$x_{\rm{Sr}}=0.05$ (b), there is a difference in the oxygen content
($\delta$) between both columns, whereas for the sample
$x_{\rm{Sr}}=0.10$ (c) there is no oxygen loss. } \label{f:phases}
\end{figure}

After reaching a maximum temperature of 525~K, the sample with 10\% Sr doping was immediately cooled down to 20~K
at 2~K/min. The stabilization of the O$_{122}$-phase observed in the
heating ramp was not reverted and the sample kept this
17\% O$_{122}$-phase during the rest of the experiment, as shown in Fig.~\ref{f:phases}.
Upon cooling we observed a series of transitions, which were
successfully identified with those expected for the coexistence of
the two phases, in the same way as reported in our previous work
\cite{09Aur}. The first transition on cooling appears at 285~K and
corresponds to the distortion of the O$_{122}$-phase occurring at
the metal-insulator transition temperature \cite{07Aur2,09Aur}.
Below this temperature, the O$_{122}$-phase is expected to present a
ferrimagnetic order in a narrow temperature range as in
YBaCo$_2$O$_{5.5}$ and YBa$_{0.95}$Sr$_{0.05}$Co$_2$O$_{5.5}$
\cite{07Aur2}. It is also possible that a concommitant structural transition occurs to a monoclinic $P112/a$ space group below the metal-insulator transition temperature, as recently proposed by Malavasi \textit{et al.}~\cite{09Mal} and Padilla-Pantoja \textit{et al.}~\cite{10Pad} based on high-resolution synchrotron X-ray diffraction data. 
Our current data do not allow a refinement of the
ferrimagnetic phase, but an increase in the diffracted intensity is
observed at the $2\theta$ position where the most intense magnetic
reflection is to be expected, as shown in Fig.~\ref{f:mag-int}. The figure
presents the intensity of the most intense magnetic peak
(corresponding to the families $\{111\}_{\rm{ferri}}$ and
$\{\frac{1}{2} \frac{1}{2} 1\}_{\rm{AFM-G}}$) normalized to the most
intense nuclear reflection. The numbers in the figure correspond to
the sequence of thermal treatments applied to the sample. Number 1
corresponds to data collected on warming of the original sample.
Number 2 corresponds to data obtained on cooling after the sample
had been heated to 525~K and 3 corresponds to the subsequent warming
to RT. The curve measured before the high--temperature treatment
indicates that there is only one magnetic phase in the sample,
corresponding to the AFM-G order in the T-phase, which is consistent
with the monophasic nature of the sample at RT. However, after having
heated to 525~K, there appears a small bump in the intensity
between 220~K and 285~K which is consistent with the ferrimagnetic
order in the O-phase. The remaining 83\% T-phase in the sample
transforms to AFM-G below 200~K, in excellent agreement with the
behavior described in our previous work \cite{09Aur}.

\begin{figure}[ptb]
\centering
\includegraphics[angle=0,width=0.6\linewidth]{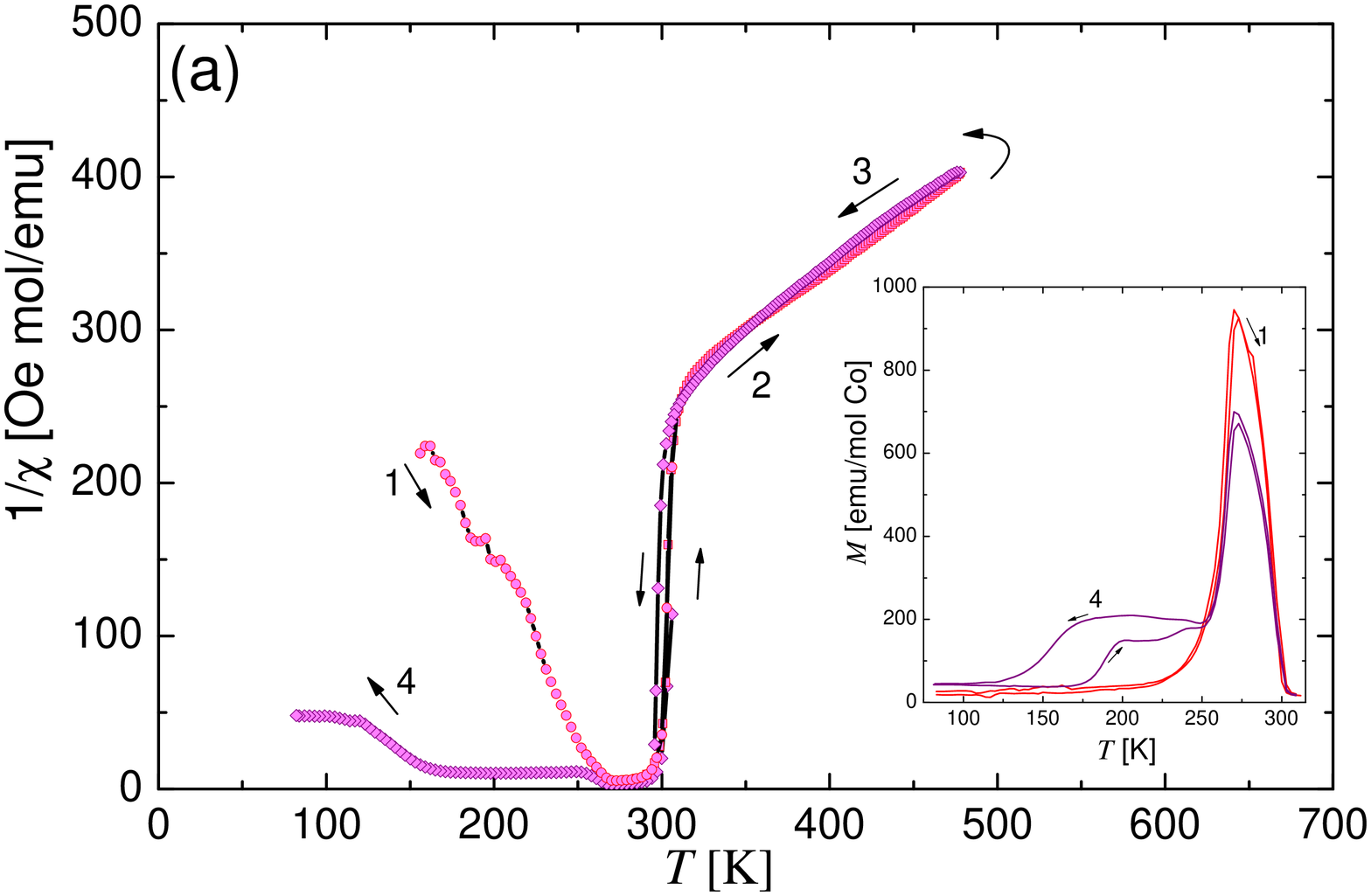}
\includegraphics[angle=0,width=0.6\linewidth]{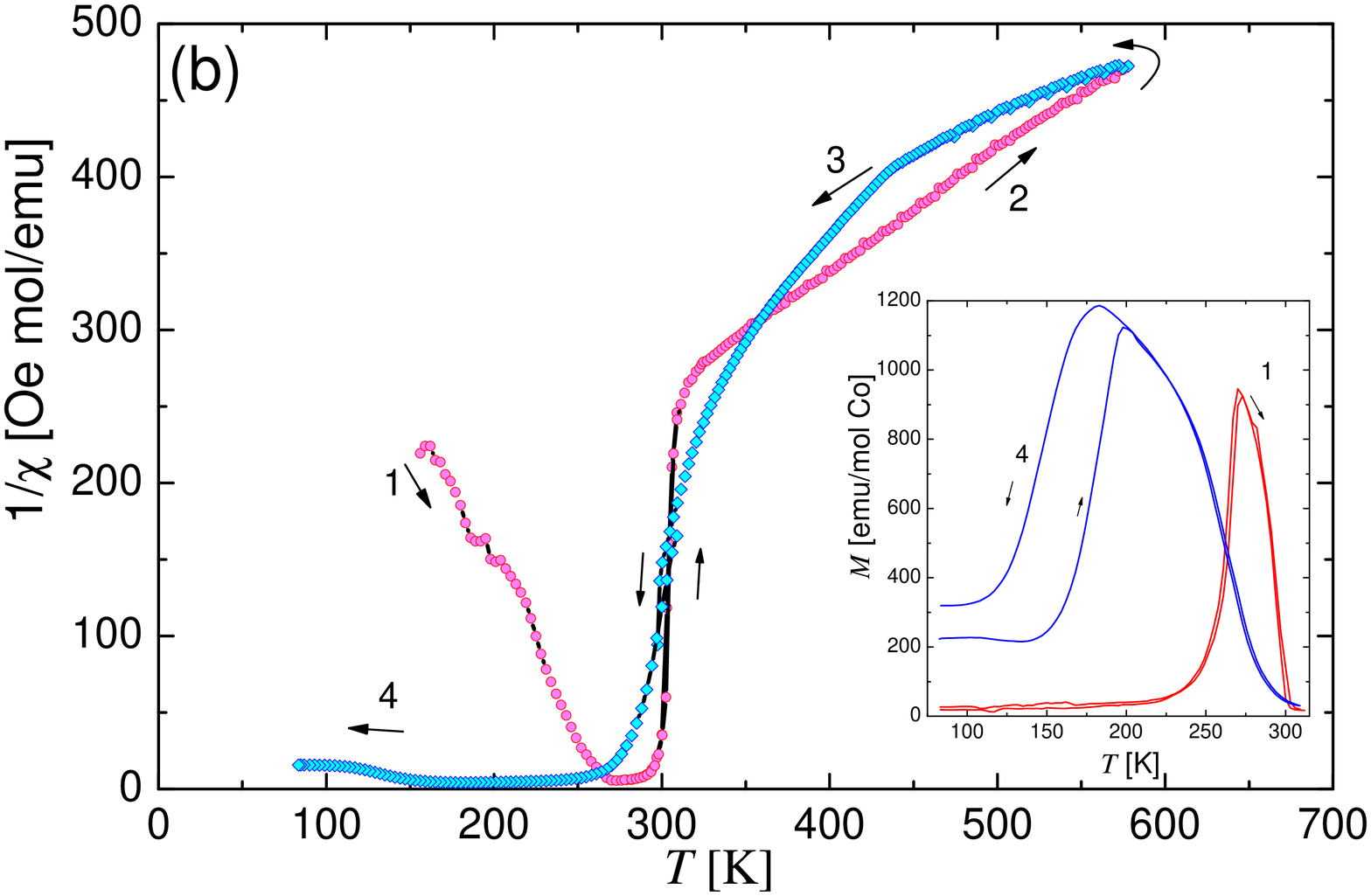}
\includegraphics[angle=0,width=0.6\linewidth]{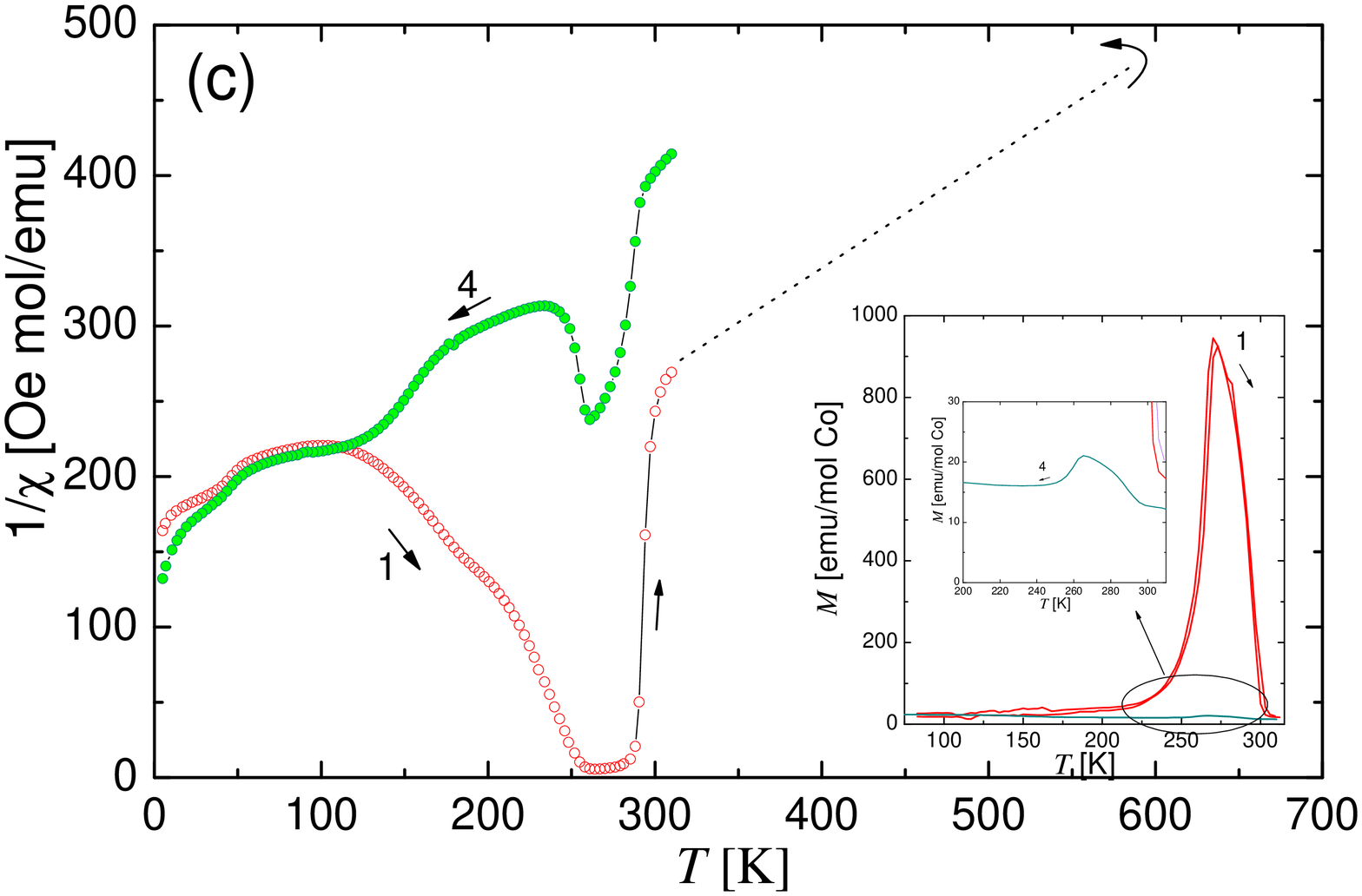}
\caption{Evolution of the inverse susceptibility at 5~kOe in
YBaCo$_{2}$O$_{5.5}$ under different thermal treatments. (a) The
as-synthesized sample was heated in a Faraday balance up to 473~K
under vacuum and slowly cooled. (b) The as-synthesized sample was
heated in a Faraday balance under vacuum up to 573~K and slowly
cooled. (c) The as-synthesized sample was heated in the
diffracometer D20 at ILL up to 570~K and slowly cooled. The inverse
susceptibility was subsequently measured in a SQUID magnetometer. In
all cases, the inset shows the low-temperature magnetization before
and after the thermal treatments.} \label{f:invchi}
\end{figure}

The thermal effects on the magnetic behavior of the parent compound
are also reflected on the susceptibility measurements.
Figure~\ref{f:invchi} shows the temperature dependence of the
inverse susceptibility of the parent compound during and after three
different thermal treatments. In panel (a) the original sample was
warmed to 473~K and immediately cooled to RT in the Faraday balance.
The inverse susceptibility data on warming and cooling do not show
any change above 300~K, but when the low-temperature magnetization
was measured again after the experiment, the resulting curve was
different. The low field magnetization (inset in
Fig.~\ref{f:invchi}(a)) suggests the presence of a second magnetic
phase. Given that we do not expect any significant oxygen loss when
heating to 473~K, this result points to a rearrangement of oxygen
vacancies and again highlights the competition of phases. It is
clear that the magnetization is the most sensitive property even to
slight variations in the amount and degree of order of oxygen atoms.
Panel (b) shows the results of a similar experiment, but this time
the sample was heated up to 573~K in the Faraday balance. As we have
seen in our neutron diffraction experiments, at such temperature we
do expect some oxygen loss. Effectively, although the sample was
immediately cooled, the inverse susceptibility curve measured on
cooling is different to the one measured on warming. The
magnetization curve also changes dramatically after the thermal
treatment, and indicates a new $\delta$ value of $\thicksim 0.40$ to
0.45 \cite{01Aka} which agrees very well with the loss of mass of
our sample. Moreover, the RT symmetry results
tetragonal instead of orthorhombic as determined by X-ray
diffraction (data not shown) in line with the work by Akahoshi~\cite{01Aka}. In the inset of
Fig.~\ref{f:invchi}(c) we present the magnetization curves collected
before and after the experiment at ILL, where the sample was heated
to 570~K and kept there for 30 minutes before cooling. The
magnetization curve agrees well with data from \cite{01Aka} for the
refined $\delta$ value of 0.24.

\begin{figure}[tb]
\centering
\includegraphics[angle=-90,width=\linewidth]{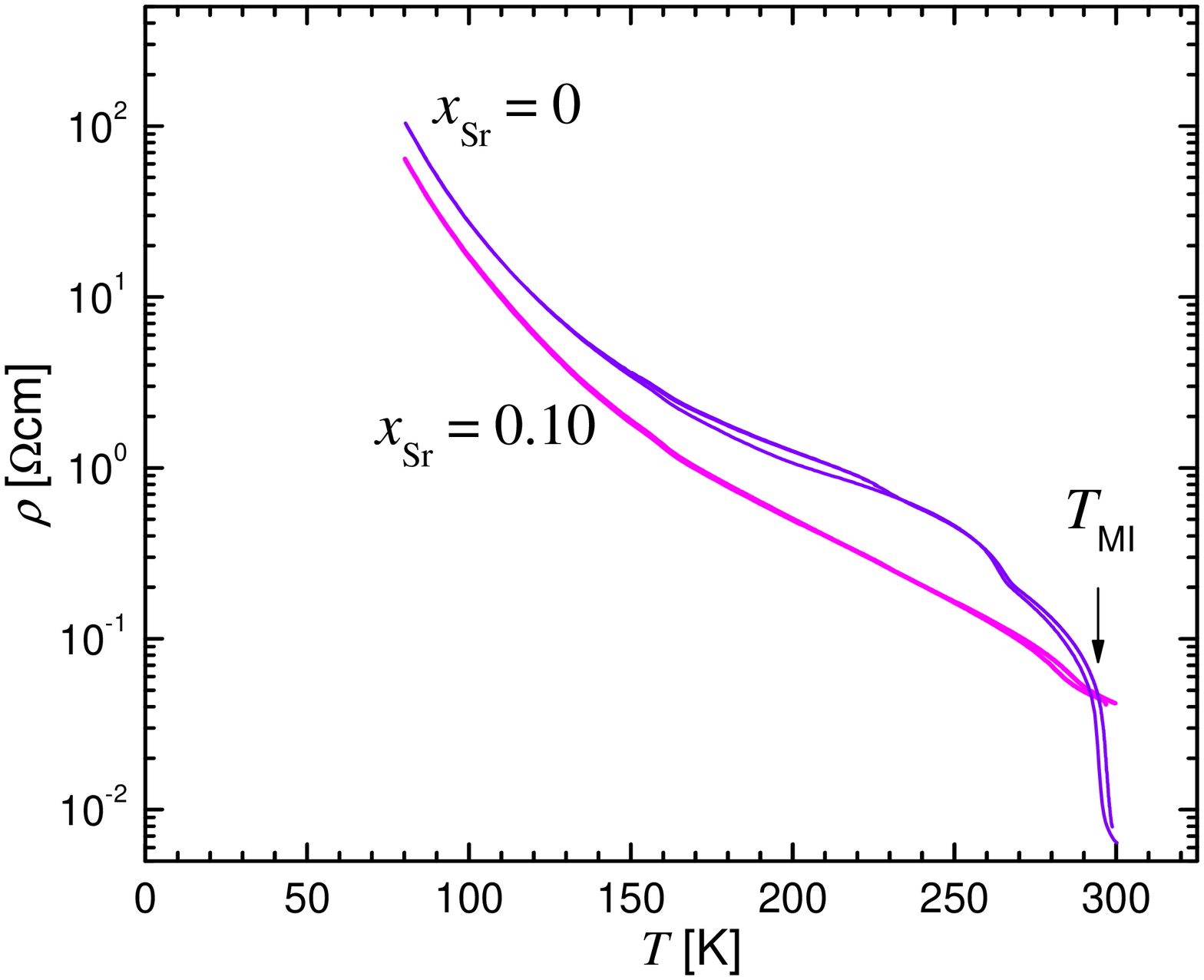}
\caption{Resistivity curves for samples with $x_{\rm{Sr}}=0$ and
$x_{\rm{Sr}}=0.10$ collected on cooling and warming in the range
80~K$<T<300$~K.} \label{f:resist}
\end{figure}

Finally, we include here a brief comment on transport properties. It has been shown that the metal-insulator transition is only observed in the O-phase, \emph{i.e.}, for $\delta$ values very
close to 0.5~\cite{05Tas}. The T-phase, on the other hand, presents
a higher resistivity above RT and behaves as a
semiconductor. To illustrate this, we present in Fig.~\ref{f:resist} the resistivity
curves under zero-applied magnetic field collected for samples with
$x_{\rm{Sr}}=0$ and $x_{\rm{Sr}}=0.10$ with $\delta= 0.47$, for the samples before been heated to high-temperatures. It must be emphasized that even though this is called a
metal-insulator transition, the behavior above RT of
the O-phase is not actually metallic, but also semiconductor-like
with a lower activation energy. On the other hand, it has been shown
for GdBaCo$_2$O$_{5+\delta}$ that the high--temperature T-phase
is metallic~\cite{08Tar} so further work on transport
properties is in progress to study the role of vacancies.

To sum up, the series of phases and transformations observed in our whole set
of experiments for the Sr-substituted samples are summarized in
Fig.~\ref{f:phases}. The column on the left represents the evolution on
warming, while the column on the right corresponds to the subsequent
cooling. Once again, it results striking how complex the phase
diagram is for $\delta \backsimeq 0.5$ when compared to lower
$\delta$ values. However, this complexity is quite consistent among
the different samples and the phase separation scenario is clearly
confirmed. For the O-phase to stabilize in this system, it is
necessary to get very close to $\delta= 0.5$: a value of $\delta=
0.45$ already presents tetragonal symmetry. The cation disorder at
the Ba site also favors the T-phase. On the contrary, an increase in
temperature favors the stabilization of the O-phase probably due to
oxygen mobility, but the sample needs to be heated in an adequate
atmosphere to prevent oxygen loss, otherwise it will transform to
the T-phase. As we have shown above, these transformations may have profound implicancies in the transport properties as they certainly have them on the magnetic behaviour, therefore the importance of building a complete phase diagram to finally correlate all these variables. 

\section{Conclusions}

The layered cobaltites \emph{R}BaCo$_{2}$O$_{5+\delta}$ have
received increasing attention in the past years, and there is still
much to be studied to understand and systematize all the variables
involved in their physical properties. In this paper, a study has been presented of the high--temperature stability and structure of the
two major phases in the $\delta=0.5$ system, the vacancy-ordered
(orthorhombic) 122-phase and the vacancy-disordered (tetragonal)
112-phase. By performing \emph{in-situ} thermal treatments in the
D20 diffractometer we could follow structural and magnetic
transitions in real time. Further structural analysis as a function of temperature will be reported in a separate article. In the present work, we have focused on the combined effect of temperature and
dopant concentration, as well as that of different oxygen
contents as a result of the experimental conditions, to obtain a mapping of these variables over the system's phase diagram. We observe that the range of stability of the O-phase lies very close to $\delta=
0.5$ and $x_{\rm{Sr}}=0$: the cation disorder at the Ba site favors
the T-phase, as well as the vacancy disordering due to oxygen loss
at high temperature. On the contrary, an increase in temperature
could favor the stabilization of the O-phase in a case in which
vacancies were initially disordered, but oxygen loss must
imperatively be prevented otherwise the system will transform to the
T-phase. The fact that samples with both $x_{\rm{Sr}}=0$ and $x_{\rm{Sr}}=0.05$ show the same final
$\delta$ value after a similar thermal treatment confirms our initial assumption that the
substitution with Sr (or Ca) has no influence on the oxygen
incorporation or desorption, but only the synthesis conditions and
thermal history do. The landscape of phases outlined
in the present work may be useful for the design of compounds with tuned physical properties. The study of these fascinating cobalt oxides is continuously in progress and further work is necessary to clarify the intimate relation between structure, transport and magnetism.

\ack

This work is part of a research project supported by Agencia
Nacional de Promoci\'{o}n Cient\'{\i }fica y Tecnol\'{o}gica
(Argentina), under grant PICT 2004-21372 and by SECTyP, Universidad
Nacional de Cuyo. We particularly acknowledge ILL and its staff for
the beamtime allocation and technical assistance.


\section*{References}
\bibliography{/home/gaurelio/Documentos/COBALTITAS/cobaltitas-biblio,/home/gaurelio/Documentos/COBALTITAS/cobaltitas-misc}
\bibliographystyle{personal-style}

\end{document}